%% file: ssgam_jpipe.tex
\documentclass[12pt]{article}

\usepackage[margin=1in]{geometry}
\usepackage{amsmath}
\usepackage{amssymb}
\usepackage{booktabs}
\usepackage{multirow}
\usepackage{graphicx}
\usepackage{caption}
\usepackage{subcaption}
\usepackage{float}
\usepackage{setspace}
\usepackage{authblk}
\usepackage[colorlinks=true,citecolor=blue,linkcolor=blue,urlcolor=blue]{hyperref}
\usepackage[style=authoryear,backend=biber,natbib=true,maxcitenames=2,%
            maxbibnames=99,giveninits=true,uniquename=false,uniquelist=false,%
            dashed=false,doi=true,isbn=false,url=false]{biblatex}
\title{Which Text Matters: Handling Model Uncertainty in Text-as-Data
  Regressions, with an Application to Central Bank Communication}

\author[1]{Martin Feldkircher}
\author[2]{Bernd Prostmaier}
\author[2]{Paul Hofmarcher\thanks{Corresponding author:
  \href{mailto:paul.hofmarcher@plus.ac.at}{paul.hofmarcher@plus.ac.at}.}}
\author[3]{Pierre L.\ Siklos}

\affil[1]{Vienna School of International Studies (DA) and Oesterreichische
  Nationalbank (OeNB), Vienna, Austria}
\affil[2]{University of Salzburg, Salzburg, Austria}
\affil[3]{Wilfrid Laurier University and Balsillie School of International
  Affairs, Waterloo, Canada}

\date{This version: \today}

\begin{document}

\maketitle

\begin{abstract}
\noindent
Social scientists routinely convert text into covariates like topic shares,
ideal points or sentiment scores, and insert them into statistical models of
interest. This two-step design is convenient, but it is rarely guided by
theory: little tells the analyst which text-based covariates to construct or
include. Each approach embodies contestable measurement choices, and competing
operationalizations of the same construct can support different conclusions,
so a single research question gives rise to a menu of competing regression
models. Which text constructed covariates belong in a model, and in what functional form, is
thus a problem of \emph{model uncertainty}. We propose spike-and-slab
generalized additive models (SSGAMs), which jointly select covariates and
determine whether their effects are linear or nonlinear. We illustrate the
approach with the communication of a powerful unelected institution, asking
which of 45 competing text indicators from 224 monthly European Central Bank
press conferences, spanning late 2002 to mid-2025, track professional
inflation expectations. A sparse set is selected; widely used alternatives,
and every tested nonlinear effect, carry no incremental information.

\end{abstract}

\section{Introduction}
\label{sec:intro}

Text has become one of political science's most important data sources. Party
manifestos, parliamentary debates, court opinions, social media posts, and the
statements of bureaucratic institutions are routinely converted into numerical
measures and analyzed with the discipline's standard statistical toolkit
\citep{grimmer2013,wilkerson2017,grimmer2022}. In the typical research design,
the text itself is not the quantity of interest. Instead, an upstream procedure
compresses it into one or more \emph{covariates}---topic proportions from a
topic model \citep{roberts2014}, policy positions or ideal points estimated
from word counts \citep{laver2003,slapin2008,vafa2020}, sentiment scores from a
dictionary or a language model \citep{young2012}, readability and
sophistication indices \citep{spirling2016,benoit2019}---which then enter a
downstream regression alongside conventional variables
\citep{fong2021,knox2022}.

This two-step design has an underappreciated consequence: the covariates are
not given, they are \emph{constructed}. Each text-based measure embodies a chain of
researcher choices---the preprocessing pipeline, the dictionary or embedding,
the model class, the aggregation rule---and these choices are consequential.
\citet{denny2018} show that seemingly innocuous preprocessing decisions can
alter the conclusions of unsupervised text analyses; \citet{rodriguez2022}
show how to assess which embedding choices matter for downstream validity;
\citet{young2012} find that competing sentiment dictionaries disagree
substantially on the same political texts. As a result,
for any latent construct of interest---tone, complexity, position---the
researcher typically faces not one measure but a menu of competing, correlated,
yet conceptually distinct operationalizations. Choosing one of them, and
entering it linearly, is the analytical equivalent of walking a single path
through a garden of forking paths \citep{gelman2014}: the reported model is one
draw from a large space of defensible specifications
\citep{steegen2016,simonsohn2020}.

We argue that this situation is best understood as a problem of \emph{model
uncertainty}, and that it admits a principled, fully probabilistic solution.
Political methodology has long recognized that uncertainty about which
covariates belong in a model should be reflected in inference rather than
suppressed by a single specification choice: \citet{bartels1997} introduced
model averaging to political science, and \citet{montgomery2010} and
\citet{montgomery2012} established Bayesian model averaging (BMA) and its
ensemble variants as practical tools for the discipline. A parallel literature
in economics has used BMA extensively to adjudicate between competing
predictors when theory is silent
\citep{salaimartin2004,hofmarcher2018,steel2020}, including work on prior
choices for model spaces with interactions \citep{ZeugnerFeldkircher,moser2014}, on the joint
inclusion patterns of correlated regressors
\citep{crespo2016,hofmarcher2015}, and on forecasting applications
\citep{Feldkircher2012,hofmarchergruen2020}. Standard BMA, however, has a structural blind
spot that is particularly damaging for text-based covariates: it averages over
\emph{which} variables enter, but forces every included variable to enter
\emph{linearly}. Whether a text feature operates linearly is itself an open
empirical question---there are good theoretical reasons to expect saturation
and non-monotonicities in how audiences process language
\citep{gibson1998,mikk2008,lewis2016,benoit2019}---and imposing linearity when the truth is
nonlinear biases estimates and can mask genuine effects.

This paper imports and adapts a framework that resolves both layers of
uncertainty simultaneously: \emph{spike-and-slab generalized additive models}
\citep[SSGAMs;][]{SSGAM,scheipl2012}. SSGAMs place a spike-and-slab prior on
groups of basis coefficients in an additive model, so that the posterior
delivers, for every candidate covariate, (i) a marginal posterior inclusion
probability---the probability that the covariate belongs in the model at
all---and (ii) a data-driven decomposition of its effect into a linear
component and a nonlinear deviation. The framework thus generalizes BMA to the
semi-parametric setting \citep{malsiner2019}, inheriting its coherent
uncertainty quantification while relaxing its most restrictive assumption. For
applied text-as-data research the payoff is direct: instead of pre-committing
to one operationalization and one functional form, the researcher can enter
the full menu of candidate text measures and let the posterior adjudicate.

We illustrate the approach with an institution for which the stakes of
language are unusually high: the European Central Bank. Central banks are
powerful, deliberately insulated, unelected institutions; their legitimacy
rests not on electoral authorization but on mandates, performance, and---
crucially---public justification \citep{tucker2018,adolph2013,baerg2025}. Since the words of a
central bank are themselves a policy instrument \citep{blinder2008}, and since
its statements are drafted, negotiated, and delivered with extraordinary care,
central bank communication has become a canonical corpus for political
science research on non-majoritarian institutions
\citep[e.g.,][]{schonhardt2013,baerglowe2020,baerg2020,moschella2020,fraccaroli2022}.
We ask a question that is substantively important and methodologically
paradigmatic: \emph{which} dimensions of the ECB's press-conference
language---sentiment, readability, syntactic and lexical complexity,
length---track the inflation expectations of professional forecasters, the
audience most directly targeted by monetary communication, and in what
functional form? The question is paradigmatic because the menu problem is
extreme here: tone alone can be measured with a general-purpose lexicon, a
domain-specific financial dictionary \citep{loughran2011}, or a transformer
model; readability with several competing indices; and every measure can be
computed separately for the scripted introductory statement and the
unscripted Q\&A exchange with journalists.

Applying SSGAMs to a comprehensive menu of text indicators---competing
sentiment lexicons, readability indices, syntactic and lexical complexity
measures, and length measures, computed separately for the scripted
introductory statement, the full Q\&A, and the president's answers within the
Q\&A---from 224 monthly ECB press conferences (December 2002--June 2025)
matched to Consensus Economics professional forecasts, we obtain three sets
of results. First, the posterior concentrates on a sparse set of indicators
from the scripted statement: of five tone-related measures spanning
affective valence and communicated policy stance, exactly one of each
construct is selected---the ECB-specific Picault--Renault monetary-policy
score and the general-purpose Syuzhet lexicon---while the finance-specific
Loughran--McDonald dictionary and the transformer-based FinBERT score are
firmly excluded. Second, functional form is tested, not assumed, and the
test comes back negative: every candidate measure enters with both a linear
and a nonlinear component, and for every complexity and readability
indicator---including the Flesch--Kincaid grade level, the closest
contender---the nonlinear component fails to clear the inclusion threshold.
A framework that only ever fit linear terms could not distinguish this
calibrated null from the alternative that nonlinearity was simply never on
the table. Third, the model is as informative about what does \emph{not}
matter: apart from one at-most-suggestive term in the president's unscripted
answers, all Q\&A-based indicators are uninformative, and the
cross-sectional dispersion of forecasts is associated with no text measure
at all. Had we pre-selected any single indicator---as is standard
practice---our substantive conclusions would have depended heavily on which
one.

The paper proceeds as follows. Section~\ref{sec:uncertainty} locates the
problem in the text-as-data literature and in the literature on model
uncertainty. Section~\ref{sec:method} presents the SSGAM framework.
Section~\ref{sec:application} introduces the application, data, and
specification. Section~\ref{sec:results} reports the results.
Section~\ref{sec:practice} distils guidance for applied researchers, and
Section~\ref{sec:conclusion} concludes.

\section{Text-based covariates and two layers of model uncertainty}
\label{sec:uncertainty}

\subsection{From text to covariates}

The text-as-data workflow almost always ends in a familiar place: a
rectangular dataset in which documents (or document--period aggregates) are
rows and text-derived measures are columns, ready for regression
\citep{grimmer2022}. The measures fall into a small number of families.
\emph{Content} measures summarize what a text is about, most prominently
topic proportions from probabilistic topic models \citep{roberts2014,
roberts2019}. \emph{Position} measures locate speakers in a latent space:
supervised scaling \citep{laver2003}, unsupervised scaling \citep{slapin2008},
ideal points estimated from behavior and text \citep{barbera2015,vafa2020}.
\emph{Tone} measures quantify emotion or polarity, via dictionaries
\citep{young2012,loughran2011}, lexicons such as Syuzhet \citep{Syuzhet,
kim2022syuzhet}, or contextual language models built on BERT
\citep{devlin2018}.\footnote{We use \emph{dictionary} to refer to resources
that assign words to discrete categories, with scoring by counting the share
of words per category (e.g.\ Loughran--McDonald); and \emph{lexicon} to
refer to resources that assign continuous valence scores to individual words,
with scoring by weighted summation (e.g.\ Syuzhet/NRC). The distinction
matters because a lexicon can distinguish mildly from strongly negative
words, while a dictionary cannot.}
\emph{Style} measures capture how something is said:
readability indices, sentence structure, and lexical sophistication
\citep{spirling2016,benoit2019,loughran2014}.
A distinct approach replaces grammatical features with a domain-specific
jargon lexicon to measure \emph{conceptual} complexity: the density and
topical breadth of technical vocabulary in a text, capturing how demanding
its content is rather than how it is grammatically packaged
\citep{mcmahon2026}.

Each family contains multiple competing operationalizations, and the
differences between them are not innocuous. Within tone, general-purpose and
domain-specific dictionaries disagree on political and financial text
\citep{young2012,loughran2011}; within style, readability formulas weight
sentence length and word length differently, while syntactic complexity is a
distinct construct again \citep{lu2010}; within content and position,
preprocessing and embedding choices propagate into the estimates
\citep{denny2018,rodriguez2022}. The upshot is that the ``text-based
covariate'' a researcher adds to a model is one member of an equivalence class
of plausible measures---correlated, but not interchangeable.

\subsection{Measurement multiplicity is model uncertainty}

How does applied work handle this multiplicity? Overwhelmingly, by choosing.
A single indicator (or a small ad hoc set) is selected before estimation, on
grounds of convention, availability, or prior literature; the regression then
proceeds conditional on that choice, and the reported standard errors ignore
it. The problem is well understood in the abstract: conditioning on one of
many defensible specifications understates uncertainty and invites---even
without any intent---the selective reporting dynamics described by
\citet{gelman2014}. The methodological response to date has been descriptive:
estimate many specifications and display the distribution of estimates, as in
multiverse analysis \citep{steegen2016} and specification-curve analysis
\citep{simonsohn2020}. These tools are valuable for transparency, but they
stop short of inference: they show that conclusions vary across the
specification space without saying which specifications the data support.

Bayesian model averaging supplies exactly that missing layer.
Given a set of $K$ candidate covariates, BMA treats each of the $2^K$ subsets
as a model, computes posterior model probabilities, and averages quantities of
interest over the model space; marginal posterior inclusion probabilities
summarize the evidence that a given covariate belongs in the model at all
\citep{bartels1997,montgomery2010,steel2020}. In economics, BMA has become the
workhorse for adjudicating among dozens of competing predictors when theory
does not pin down the specification \citep{salaimartin2004}, and a
substantial literature has refined its machinery: priors over model spaces
\citep{moser2014}, the treatment of correlated regressors
\citep{hofmarcher2015}, and measures of joint inclusion that reveal whether
predictors are substitutes or complements in the model space
\citep{hofmarcher2018,crespo2016}. Text-based covariates are a natural---and
so far largely untapped---domain for this machinery, precisely because
measurement multiplicity guarantees a large space of correlated candidates.

\subsection{The functional-form blind spot}

BMA, however, resolves only the first layer of uncertainty. Its second,
usually implicit, assumption is that every included covariate acts linearly on
the outcome. For text-based measures this assumption is fragile. Cognitive
and linguistic research gives reasons to expect non-monotonic and saturating
relationships: word length is a correlate of conceptual complexity
\citep{lewis2016}, and the processing cost of syntactically complex language
rises with the difficulty of the audience's task, suggesting threshold rather
than proportional effects \citep{benoit2019}. Empirically, forcing a
nonlinear relationship into a linear term biases the coefficient towards zero
and can flip model-selection verdicts: a covariate whose effect is
hump-shaped may be discarded by a linear selection procedure altogether.
Functional form is thus not a nuisance detail but a second dimension of the
model space---and it should be selected over, and averaged over, with the
same discipline as variable inclusion. This is precisely what spike-and-slab
generalized additive models provide.

\section{Spike-and-slab generalized additive models}
\label{sec:method}

\subsection{The additive model and the linear--nonlinear decomposition}

Let $\mathbf{y}\in\mathbb{R}^n$ denote the outcome and $x_1,\ldots,x_K$ the
candidate covariates---in our application, $K=45$ text indicators plus
categorical controls. The base model is additive:
\begin{equation}
  y_i = \alpha + \sum_{k=1}^{K} f_k(x_{ik}) + \varepsilon_i, \qquad
  \varepsilon_i \overset{iid}{\sim} \mathcal{N}(0,\sigma^2),
  \label{eq:gam}
\end{equation}
where each $f_k$ is an unknown smooth function. Each smooth is represented by
a penalized B-spline basis \citep{eilers1996} and---following
\citet{scheipl2012}---orthogonally decomposed into a strictly linear part and
a nonlinear deviation:
\begin{equation}
  f_k(x_k) = x_k\beta_k + \mathbf{Z}_k\boldsymbol{\theta}_k,
  \label{eq:decomp}
\end{equation}
where the column space of $\mathbf{Z}_k$ is orthogonal to the intercept and to
$x_k$. The term $x_k\beta_k$ captures the best linear approximation of
$f_k$; $\mathbf{Z}_k\boldsymbol{\theta}_k$ captures residual curvature. The
full model is
\begin{equation}
  \mathbf{y} = \alpha\mathbf{1}_n +
  \sum_{k=1}^{K}\bigl(x_k\beta_k + \mathbf{Z}_k\boldsymbol{\theta}_k\bigr)
  + \boldsymbol{\varepsilon}, \qquad
  \boldsymbol{\varepsilon}\sim\mathcal{N}(\mathbf{0},\sigma^2\mathbf{I}).
  \label{eq:full}
\end{equation}
Because the linear part and the nonlinear deviation of each covariate carry
separate inclusion indicators, the posterior distinguishes three regimes for
every covariate: \emph{exclusion} (neither component selected), \emph{linear
inclusion} (only $x_k\beta_k$ selected), and \emph{nonlinear inclusion} (the
smooth deviation selected). Categorical covariates enter as factor terms with
their own indicators. Variable selection and functional-form selection are
thus decided \emph{jointly}, from the data, within a single posterior---rather
than sequentially by the researcher.

\subsection{The peNMIG prior}

Selection is implemented through the \emph{parameter-expanded
normal-mixture-of-inverse-Gammas} (peNMIG) prior of \citet{scheipl2012}. Each
coefficient block $\boldsymbol{\beta}_k$ (a scalar for linear terms, a vector
for smooth and factor terms) is reparameterized as
$\boldsymbol{\beta}_k=\alpha_k\boldsymbol{\xi}_k$, where the scalar $\alpha_k$
carries the overall magnitude of the term and $\boldsymbol{\xi}_k$ distributes
it across basis functions. A spike-and-slab mixture on the block variance
$\tau^2_k$ governs inclusion:
\begin{equation}
  \tau_k^2 \;\Big|\; \gamma_k \;\sim\;
  \gamma_k \cdot \tau^2_{\mathrm{slab}}
  + (1-\gamma_k)\cdot v_0\,\tau^2_{\mathrm{slab}},
  \label{eq:prior_mix}
\end{equation}
with spike factor $v_0 = 2.5\times10^{-4}$,
$\tau^2_{\mathrm{slab}}\sim\mathrm{IG}(5,25)$, and inclusion weight
$w\sim\mathrm{Beta}(1,1)$, implying a prior expected inclusion rate of 50\%
per term. The spike shrinks excluded blocks to approximately zero while the
slab leaves active blocks essentially unconstrained---the classical
spike-and-slab logic \citep{mitchellbeauchamp1988} lifted to groups of basis
coefficients. \citet{scheipl2012} show that this prior reliably recovers
sparse truths and correctly discriminates linear from nonlinear effects in
finite samples; \citet{malsiner2019} demonstrate its performance on economic
data with many heterogeneous predictors, where it compares favorably to
linear shrinkage and selection benchmarks.

\subsection{Posterior summaries}

We summarize the posterior with two complementary quantities per term. The
\emph{marginal posterior inclusion probability} (PIP)
$P(\gamma_k=1\mid\mathbf{y})$,
the posterior mean of the inclusion indicator, quantifies the evidence that
term $k$ belongs in the model; it is the direct analogue of the BMA inclusion
probability familiar to political scientists \citep{montgomery2010}. Because
the $\mathrm{Beta}(1,1)$ prior on the inclusion weight implies a prior
inclusion probability of one half for every term, the posterior odds
$P(\gamma_k=1)/\{1-P(\gamma_k=1)\}$ can be read directly as the weight of
evidence for including term $k$: probabilities near $0.5$ indicate that the
data have not moved the prior, and probabilities \emph{below} $0.5$ indicate
evidence for exclusion. The
\emph{term importance} $\hat\pi_k$ measures how much of the model's fitted
systematic variation is attributable to term $k$: writing
$\hat{\boldsymbol{\eta}}_k$ for the posterior-mean contribution of term $k$ to
the linear predictor and
$\hat{\boldsymbol{\eta}}=\sum_j\hat{\boldsymbol{\eta}}_j$ for the total
posterior-mean linear predictor net of the intercept,
\begin{equation}
  \hat\pi_k \;=\;
  \frac{\hat{\boldsymbol{\eta}}_k^{\prime}\hat{\boldsymbol{\eta}}}
       {\hat{\boldsymbol{\eta}}^{\prime}\hat{\boldsymbol{\eta}}},
  \qquad \textstyle\sum_k \hat\pi_k = 1,
  \label{eq:pi}
\end{equation}
so that the importances of all terms decompose the squared norm of the fitted
linear predictor \citep{SSGAM}. Because contributions of correlated terms can
partially offset, $\hat\pi_k$ is a relative importance weight; we interpret
its magnitude, and we characterize the \emph{shape} of selected effects
through posterior partial-effect plots rather than through scalar summaries.
Together, $P(\gamma_k=1)$ and $\hat\pi_k$ answer the two questions an applied
researcher asks of every candidate covariate: does it belong in the model,
and how much of the explained variation does it account for? In addition, the
posterior over entire models is available, so one can report the posterior
probability of the \emph{null} model as a global check of whether the
covariate set explains anything at all.

\subsection{Implementation}

Estimation uses the \texttt{spikeSlabGAM} package in \textsf{R}
\citep{SSGAM}, which implements blockwise Gibbs sampling for the peNMIG
prior. We run 5 parallel chains, each for 10\,000 iterations with a burn-in
of 5\,000 and thinning by 5, yielding 1\,000 retained draws per chain
(5\,000 in total). Convergence is monitored via the Gelman--Rubin potential scale reduction
factor $\hat{R}$ \citep{gelman1992}, with values below 1.1 taken to
indicate convergence. All continuous covariates are winsorized at the
1st and 99th percentiles before estimation.

\section{Application: the justificatory language of an unelected institution}
\label{sec:application}

\subsection{Why central bank communication?}

Central banks are the paradigmatic case of delegated, non-majoritarian
power. Their independence is
constitutionally entrenched precisely so that elected politicians cannot
easily override them \citep{adolph2013,bodea2015}, which makes public
justification---rather than electoral sanction---their principal channel of
accountability \citep{tucker2018,braun2016,goodhart2018}.
The ECB is an extreme case even within this class: it answers to no single
government, serves twenty national publics, and has repeatedly been pushed---
through the sovereign debt crisis, quantitative easing, and the pandemic---into
policies whose distributive consequences are politically contested. How such
an institution argues for its policy, and whether its arguments actually
steer the beliefs of its audiences, is therefore a first-order question for
the politics of delegation, not merely for monetary economics.

A substantial literature has established that central bank words move
markets and expectations \citep[for a survey, see][]{moessner2017}:
announcements and their tone move asset prices
\citep{altavilla2019,ehrmann2007,gorodnichenko2023,kanelis2026}, tone shapes
professional forecasts \citep{hubert2021}, communication affects household
expectations under some conditions \citep{coibion2022,feldkircher2025}, and
the linguistic complexity of ECB press conferences influences media engagement
\citep{ferrara2022} and financial market trading \citep{hayo2022}. Textual
methods have likewise entered the study of monetary institutions themselves:
deliberation inside the FOMC \citep{schonhardt2013,hansen2018},
mandate differences and communication content \citep{bohl2023}, and
ideological heterogeneity across euro-area central banks
\citep{feldkircher2024}. Yet this literature shares the design weakness
described in Section~\ref{sec:uncertainty}: virtually every study pre-selects
one or two text indicators---a particular dictionary, a particular
readability formula---before estimation, and the survey literature documents
a proliferation of competing measurement approaches
\citep{blinder2008,dehaan2019,blinder2024}. The ECB press conference sharpens the problem
further because it consists of two informationally distinct segments: a
scripted introductory statement, drafted and approved by the Governing
Council, and an unscripted Q\&A session with journalists. Whether the
carefully negotiated script or the spontaneous exchange carries the
information---and on which linguistic dimension---is exactly the kind of
question that should not be settled by indicator fiat.

\subsection{Data}

We use ECB press conference transcripts matched to Consensus Economics one-year-ahead
CPI inflation forecasts from December 2002 to June 2025: 224 monthly
observations in the one-month lead ($t+1$) baseline specification and 223 in
the contemporaneous ($t$) robustness specification, which lacks an outcome
for the initial month (eleven to twelve press conferences per year until
2014 and eight per year thereafter, following the ECB's meeting calendar).
Following \citet{feldkircher2025}, we conceptualize central bank communication
as multidimensional. Our sample is restricted to monetary-policy press
conferences: we exclude eight events from the raw corpus that are not tied
to a Governing Council monetary-policy meeting---among them a strategy-review
press seminar, a euro-banknote design event, and a comprehensive-assessment
briefing\footnote{The full list, identified from press-conference titles and
excluded by document identifier, is documented in the replication
materials. One of these non-monetary-policy events shares a calendar month
with a genuine monetary-policy press conference (May 2003), so the exclusion
is applied at the event level, not the month level.}---so that thematic
variation is limited and we can focus cleanly on differences in sentiment and
linguistic structure. One monthly observation (August 2006) aggregates two
genuine monetary-policy press conferences, held three weeks apart because
that year's September meeting was moved forward to August 31; one further
monetary-policy press conference (July 2019) is present in the corpus but
excluded from the estimation sample because an incomplete transcript parse
leaves several Q\&A-answer measures missing for that event.

We compute indicators for three text sources per event: the introductory statement
(\emph{intro}), the full Q\&A session (\emph{qa}), and the ECB president's answers
within the Q\&A (\emph{qa\_ans}). \emph{Sentiment} is measured using the Syuzhet
lexicon \citep{Syuzhet,kim2022syuzhet}, the Loughran--McDonald (LM) dictionary
\citep{loughran2011}, and a transformer-based model (FinBERT;
\citealp{araci2019,huang2023}). We additionally compute
the monetary-policy and economic-environment tone scores from the domain-specific
Picault--Renault ECB lexicon \citep{picault2017}, applied to the introductory statement
and full Q\&A. \emph{Readability} is captured by Flesch Reading Ease,
Flesch--Kincaid Grade Level, and ARI. \emph{Syntactic and lexical complexity} enters
through average words per sentence, the complex-sentence ratio, average characters
per word, the type--token ratio (the number of distinct word forms divided by
the total number of words in a statement), and a conceptual complexity index
(CCI) following \citet{mcmahon2026}. Three \emph{length} indicators complete the set per text
source, yielding 45 text-based covariates in total.
See Table~\ref{tab:measures} in the appendix for definitions.

Our dependent variable is the one-year-ahead CPI inflation forecast from Consensus
Economics, transformed into a fixed-horizon measure following \citet{siklos2013}.
We denote by $\pi^{e}_{t+1}$ the cross-sectional mean of these fixed-horizon
forecasts collected in survey month $t+1$---our measure of the forecast
\emph{level}---and by $\sigma(\pi^{e}_{t+1})$ the corresponding cross-sectional
standard deviation across individual forecasters, our measure of forecast
\emph{disagreement}, which we use as a common, though
imperfect, proxy for expectation uncertainty. Matching text measures to
forecasts by calendar month is complicated by within-month timing: Consensus
Economics surveys are typically fielded around mid-month, and ECB press
conferences also fall within the second or third week of the month, so the
ordering of survey fieldwork and press conference varies across observations.
A press conference matched to the same-month survey may therefore postdate the
survey it is meant to explain, which would let expectations already held by
forecasters feed back into how the survey is coded rather than the reverse.
We resolve this by adopting the one-month lead convention as our baseline: the
survey collected in month $t+1$ serves as the outcome for the press conference
held in month $t$, with fixed-horizon weights computed on the basis of the
$t+1$ survey month. This timing ensures that the communication variable
strictly precedes the measured expectation for every observation, addressing
the potential simultaneity of the same-month match. The fixed-horizon
transformation, which interpolates across consecutive survey vintages, further
reduces sensitivity to the residual within-month timing variation. Results for
the contemporaneous ($t$) specification, in which the same-month survey serves
as the outcome, are reported as a robustness check in
Table~\ref{tab:pip_t0} in the appendix; the two specifications agree on which
text measures are selected. The covariate set further includes ECB president dummies
(Trichet, Draghi, Lagarde) and crisis-period indicators (GFC 2007--09, Euro-area
sovereign debt crisis 2010--12, COVID-19 2020--21).

\subsection{Specification}

Both outcomes---the forecast level and forecast dispersion---are modeled with
the same SSGAM specification: every text indicator enters with a linear term
and a nonlinear smooth deviation, and the presidency and crisis indicators
enter as factor terms. Priors and MCMC settings are as described in
Section~\ref{sec:method}, and all continuous covariates are winsorized at the
1st and 99th percentiles; raw and standardized preprocessing variants serve
as robustness checks.

Two design choices deserve emphasis. First, the candidate set contains no
macroeconomic fundamentals: the model asks which text measures covary with
professional expectations, not whether text moves expectations conditional on
the macroeconomic state. Second, the disturbances are treated as independent,
while both outcomes are persistent time series of overlapping fixed-horizon
forecasts. Inclusion probabilities should therefore be read as descriptive
measures of systematic covariation rather than as tests with exact error
rates; serially correlated errors or macroeconomic controls in the candidate
set are natural extensions, to which we return in
Section~\ref{sec:practice}, and a specification that adds a lagged outcome
to the candidate set is examined in Section~\ref{sec:robustness}.

\section{Results}
\label{sec:results}

The posterior delivers a sparse verdict: of the 45 candidate text measures,
only two---the Picault--Renault monetary-policy score and the Syuzhet
sentiment score, both computed from the scripted introductory statement and
both entering linearly---clear the inclusion threshold of one half for the
forecast level, and none does for forecast dispersion. No measure clears the
threshold for a nonlinear effect. Two categorical controls, the Lagarde
presidency and the COVID-19 crisis indicator, also clear the threshold and
are discussed separately below. Table~\ref{tab:pip} reports marginal
posterior inclusion probabilities (PIPs), posterior mean coefficients, and
term importances for the main terms of interest, for both outcomes, from the
winsorized specification. Throughout, we benchmark each PIP against its
prior value of one half, so that readers can translate the reported
probabilities into posterior odds.\footnote{Model-fit statistics and MCMC
convergence diagnostics for this specification are reported in
Table~\ref{tab:mcmc_diag_t1} in the appendix; the corresponding diagnostics
for the contemporaneous specification are reported in
Table~\ref{tab:mcmc_diag_t0}. All reported parameters converge comfortably
by the $\hat{R}<1.1$ benchmark, with $\hat{R}\leq1.01$ throughout.}

\begin{table}[H]
\centering
\caption{Posterior inclusion probabilities, posterior mean coefficients, and term importance}
\label{tab:pip}
\small
\input{Figures/pip_table_t1_18.tex}
\end{table}

\subsection{Forecast level: which tone measures survive a crowded menu?}

The candidate set confronts the model with five tone-related measures that
span two distinct constructs: affective \emph{valence}---the general-purpose
Syuzhet lexicon, the finance-specific Loughran--McDonald dictionary, and the
transformer-based FinBERT score---and communicated \emph{policy
stance}---the monetary-policy and economic-environment dimensions of the
ECB-specific Picault--Renault lexicon, which score dovish versus hawkish
language. The posterior concentrates on exactly one measure of each
construct, both computed from the scripted introductory statement. The
Picault--Renault monetary-policy score is the most robustly selected text
term in the system ($P(\gamma=1)=0.841$, posterior odds $5.3$;
$\hat\pi=0.18$) followed by the Syuzhet sentiment score
 ($P=0.660$, posterior odds $1.9$; $\hat\pi=0.15$). Both variables enter the specification in their linear forms only.
The positive coefficient on the Picault--Renault monetary-policy score implies
that a more hawkish introductory statement is associated with higher
professional inflation expectations---consistent with a reading in which the
ECB's hawkish signals reflect inflationary pressures that forecasters
independently assess as persistent, or, put differently, that monetary policy
is behind the curve.
The economic-environment dimension of the same lexicon is not selected
($P=0.034$), a result best understood as absorption into the retained Syuzhet
general-purpose sentiment score: both measures capture affective tone directed
at the economic outlook, and once the general-purpose lexicon enters the model
the domain-specific dimension contributes no incremental information. The
negative Syuzhet coefficient indicates that more optimistic
introductory-statement language is associated with lower professional CPI
inflation expectations, consistent with positive ECB communication reflecting
periods of price stability in which forecasters anchor expectations at lower
levels.
The remaining tone measures are firmly excluded: the
Loughran--McDonald dictionary ($P=0.029$), FinBERT ($P=0.020$), and the
economic-environment dimension of the Picault--Renault lexicon ($P=0.034$). The
pattern cuts across the categories that usually guide indicator choice:
domain-specificity per se is not what the data reward, since the discarded
measures include both a finance dictionary and a domain-adapted language
model. What the posterior retains is one measure of each \emph{construct}. The two
selected measures enter with opposite signs---the Picault--Renault score
positively and Syuzhet negatively---consistent with their capturing distinct
dimensions of the ECB's communication: policy stance versus general affective
valence. That stance and valence are nearly orthogonal in the raw data
($r=0.24$), whereas
the discarded Loughran--McDonald and FinBERT scores correlate strongly with
the retained Syuzhet lexicon ($r=0.52$ and $0.59$). The model keeps two
complementary dimensions of communication and discards the largely redundant
duplicates---an adjudication a pre-committed single-indicator design never
performs.

\begin{figure}[H]
  \centering
  \begin{subfigure}[b]{0.45\textwidth}
    \centering
    \includegraphics[width=\linewidth]{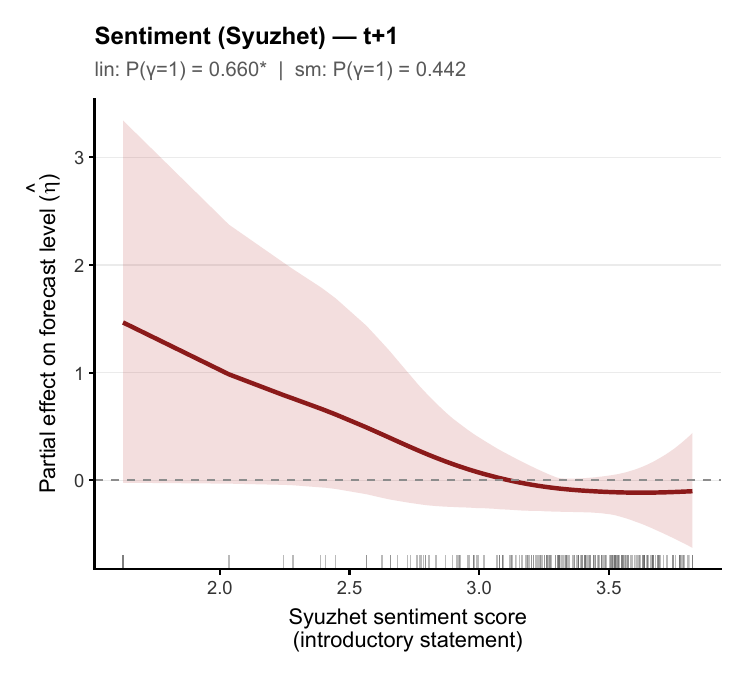}
    \subcaption{Syuzhet sentiment (\texttt{intro\_syuz})}
    \label{fig:syuz}
  \end{subfigure}
  \hfill
  \begin{subfigure}[b]{0.45\textwidth}
    \centering
    \includegraphics[width=\linewidth]{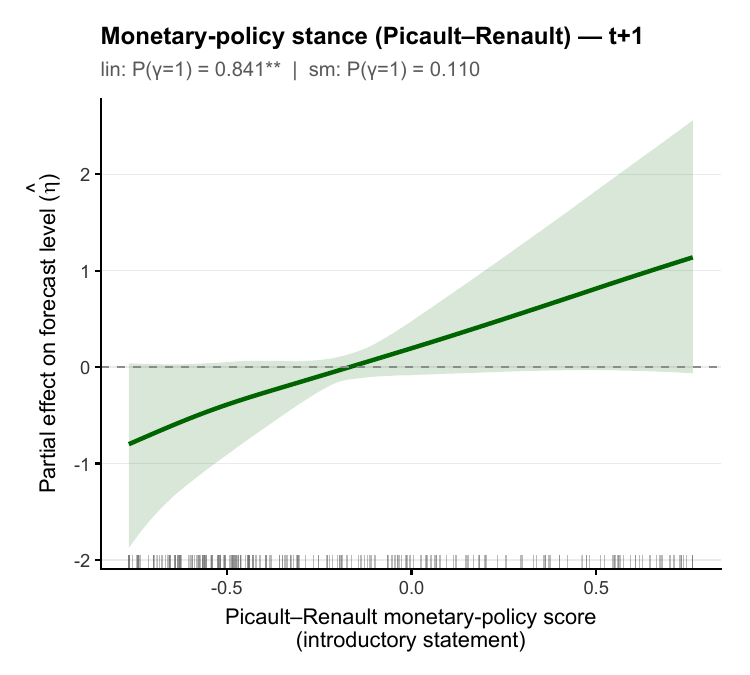}
    \subcaption{Monetary-policy stance (\texttt{intro\_mp\_pr})}
    \label{fig:mp}
  \end{subfigure}

  \medskip
  \begin{subfigure}[b]{0.45\textwidth}
    \centering
    \includegraphics[width=\linewidth]{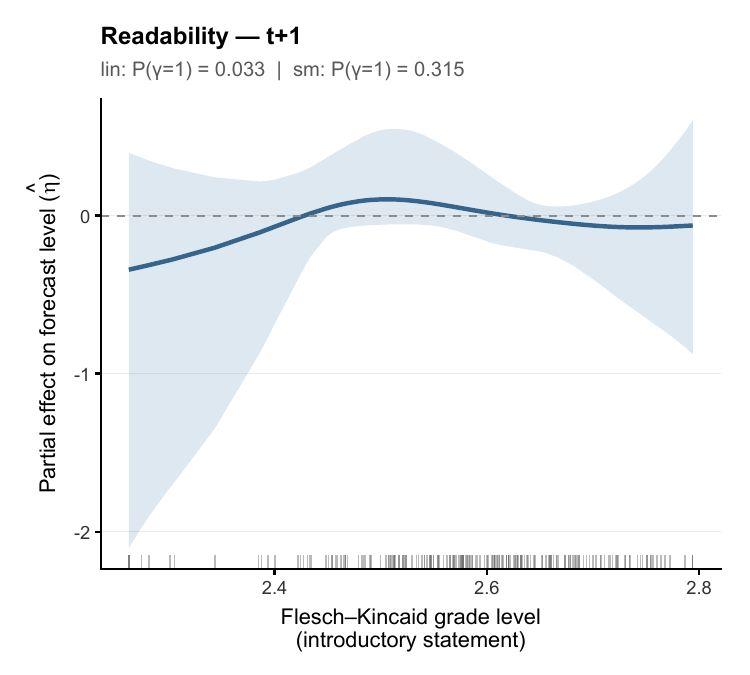}
    \subcaption{Flesch--Kincaid grade (\texttt{intro\_fk\_grade}); not selected}
    \label{fig:fkg}
  \end{subfigure}
  \caption{Posterior partial effects, one-month lead ($t+1$) specification
           (winsorized): panels (a)--(b) show the two selected
           introductory-statement tone measures; panel (c) shows the
           Flesch--Kincaid grade level as an illustrative example of a
           nonlinear term, although it is not selected}
  \label{fig:selected_tone}
  \smallskip\par
  \begin{minipage}{0.95\textwidth}
    \footnotesize\emph{Notes:} Posterior mean partial effect (solid line) and
    95\% credible band (5,000 MCMC draws, 5 chains). Rug marks show the
    predictor's empirical distribution. $t+1$ specification; winsorized; no
    macro controls.
  \end{minipage}
\end{figure}

Figure~\ref{fig:selected_tone} shows the posterior partial effect of each
selected tone measure, constructed from both its linear and nonlinear
component. For both, the effect is essentially linear, since the PIPs
attached to the nonlinear terms remain below the inclusion threshold
(Table~\ref{tab:pip}). The plot
confirms the main result: the Picault--Renault score is positively related
to one-year-ahead forecast expectations, while Syuzhet sentiment is
negatively and linearly related to them. Panel (c) adds the
Flesch--Kincaid grade level purely as an illustration of what a nonlinear
partial effect looks like within this framework: although neither its
linear nor its nonlinear component is selected here, the shape of the
estimated curve---and the width of its credible band---makes concrete what
the decomposition in equation~\eqref{eq:decomp} tests for.

Next, we investigate readability and syntactic complexity. Here the
framework's contribution is a calibrated null rather than a discovery: every
measure enters with both a linear and a nonlinear component, and none of
either kind clears the inclusion threshold. For the Flesch--Kincaid grade
level of the introductory statement, the closest contender, the data rule
out a \emph{linear} term ($P=0.033$) and offer only sub-threshold, and thus
exclusion-favoring, support for a nonlinear one ($P=0.315$, posterior odds
$0.46$). Figure~\ref{fig:selected_tone}(c) plots its estimated effect
purely as an example of what a nonlinear term looks like within this
framework, although it is not selected here: a shallow, near-flat wiggle
whose 95\% credible band contains zero across almost the entire observed
range of readability. This matters methodologically: a linear-only framework,
including standard BMA, would
have reported the same conclusion---readability does not matter---but would
have been unable to distinguish a genuine null from a nonlinear effect it
was never equipped to detect in the first place. The decomposition in
equation~\eqref{eq:decomp} tests for both and reports that neither is
present, which is a more informative statement than either a bare
insignificant linear coefficient or a design that never tested the
alternative.

The remaining complexity measures likewise show weak, sub-threshold support:
the multi-clause sentence ratio ($P=0.236$), the sentence-length smooth
($P=0.257$), and average word length ($P=0.094$). Inclusion probabilities
are statements about a variable's contribution \emph{relative to a candidate
set}, not about a variable in isolation---a further reason to enter the menu
rather than a hand-picked subset. Among the categorical controls, the Lagarde
presidency is the single most strongly selected term overall ($P=0.946$,
posterior odds $17.5$; $\hat\pi=0.50$), and the COVID-19 crisis indicator is
also selected ($P=0.623$, posterior odds $1.7$; $\hat\pi=0.03$); the Trichet,
Draghi, and the remaining crisis-period terms receive little support. We
deliberately refrain from a substantive reading of these two terms: they
partition the sample into a few long regimes that coincide with the
post-2020 inflation surge, so they are best understood as period effects
that absorb macroeconomic variation---and thereby discipline the text
terms---rather than as communication effects (see \citealp{feldkircher2024}
on heterogeneity across ECB presidencies).
Among the text-based predictors, the two selected tone measures account
for the largest shares of explained variation: the Picault--Renault
monetary-policy score contributes $\hat\pi=0.18$ and Syuzhet sentiment
$\hat\pi=0.15$.

While we do find robustly related text indicators of the scripted introductory statement, we do not find any measure from either the full Q\&A or the president's answers within it robustly related to inflation expectations. The only measure that comes close to being included is the type--token ratio of the president's answers ($P=0.429$), a measure that assesses how often words are repeated in a document, entering with a near-linear positive slope whose 95\% credible band still contains zero across most of the observed range. The monetary-policy
stance of the full Q\&A likewise remains clearly excluded ($P=0.171$), as does every
other Q\&A term. The information content of the press
conference thus resides overwhelmingly in the carefully negotiated script,
complementing the market-microstructure evidence of \citet{hayo2022} that
complex statements shift trading activity toward the Q\&A without the Q\&A
adding predictive content of its own.

For forecast disagreement the verdict is uniform: no text-based term exceeds
$P(\gamma=1)=0.02$---far below the prior inclusion probability of one half,
so the data actively favor exclusion for every term. Professional forecast dispersion, in
this sample, is not a function of press-conference language---extending the
evidence on the sources of forecast disagreement in \citet{siklos2013} to the
textual dimension. We stress the methodological value of reporting this null:
because inclusion probabilities quantify support for exclusion as well as
inclusion, the framework distinguishes ``no evidence of an effect'' from
``evidence of no effect'' in a way that a battery of insignificant
single-indicator regressions cannot.

\subsection{What a single-indicator design would have concluded}

The results double as an in-sample demonstration of the stakes of indicator
pre-selection---and rather than inferring those stakes from the joint
posterior, we run the pre-selected designs themselves. For each of the five
tone measures, and for the Flesch--Kincaid grade level entered linearly as a
conventional design would, we regress the forecast level on that indicator
alone, and again with the presidency and crisis-period dummies. Disturbances
are treated as independent, exactly as in the SSGAM specification, so the
single-indicator designs and the joint model are compared on the same
footing and the significance statements carry the same descriptive caveat
(Section~\ref{sec:application}). Table~\ref{tab:single_indicator} reports
the results.

\begin{table}[H]
\centering
\caption{Single-indicator designs --- one-month lead ($t+1$) forecast level}
\label{tab:single_indicator}
\smallskip\par
\small
\input{Figures/single_indicator_ols_22.tex}
\end{table}

The exercise shows that a single-indicator design would rarely have
concluded that communication carries no signal---it would instead have
attributed the signal to whichever indicator the researcher happened to
choose. Five of the six indicators attract a coefficient significant at the
5\% level when entered alone; the verdicts, however, do not survive contact
with each other. All three valence lexicons are individually significant,
yet the joint posterior shows that two of them owe their bivariate appeal to
their strong correlation with the retained Syuzhet score ($r=0.52$ and
$0.59$): once the full menu competes within one model, the
Loughran--McDonald and FinBERT terms are firmly excluded---and in the
single-indicator designs both already lose significance once the regime
controls enter. The linear readability term is significantly negative on its
own, a verdict the joint model does not corroborate: neither the linear nor
the nonlinear Flesch--Kincaid component clears the inclusion threshold once
the full menu of correlated text measures competes within one model
($P=0.033$ and $P=0.315$, respectively)---a bivariate association that a
design testing only one indicator at a time cannot distinguish from a
genuine readability effect. The economic-environment score is significant only with the
controls, the monetary-policy score in both variants. In sum, eight of the
twelve design variants deliver a significant text coefficient, no two
indicator choices agree on the pattern, and a researcher committed to any
single one of them would have written a different paper. The joint framework
replaces this lottery with a single posterior over the full menu.

\subsection{Robustness analysis}
\label{sec:robustness}

We probe the robustness of the selection along three dimensions: the timing
convention that links press conferences to survey vintages, the winsorization
of the continuous covariates, and the persistence of the outcome.

Under the contemporaneous convention, in which the same-month survey serves
as the outcome, the selection is essentially unchanged
(Table~\ref{tab:pip_t0} and Figure~\ref{fig:t0_partial} in the appendix).
The Picault--Renault monetary-policy score ($P(\gamma=1)=0.822$; baseline
$0.841$) and the Syuzhet score ($P=0.720$; baseline $0.660$) remain the two
selected tone terms, with unchanged signs; the functional-form null carries
over as well, with the Flesch--Kincaid nonlinear component again
sub-threshold ($P=0.281$; baseline $0.315$); and the Lagarde and COVID-19
indicators are again the dominant categorical terms ($P=0.930$ and
$0.602$). The dispersion null is, if anything, sharper: no text-based term
exceeds $P=0.01$. The type--token ratio of the president's answers, the
closest Q\&A contender under both conventions, remains sub-threshold here
too ($P=0.460$; baseline $0.429$), so the two timing conventions agree not
only on what is selected but on what is not.

Winsorizing at the 1st and 99th percentiles trims exactly the tail
episodes---crisis press conferences---in which communication may matter
most, so the selection could in principle hinge on it. We therefore
re-estimate the baseline ($t+1$) specification on non-winsorized covariates;
the resulting inclusion probabilities are reported in
Table~\ref{tab:pip_nonwinsor} in the appendix. The selection is essentially
invariant to this choice: exactly the same four terms clear the inclusion
threshold with and without winsorization---the Picault--Renault
monetary-policy score ($P(\gamma=1)=0.846$ non-winsorized versus $0.841$ in
the baseline), Syuzhet sentiment ($0.739$ versus $0.660$), and the Lagarde
and COVID-19 indicators ($0.953$ versus $0.946$ and $0.703$ versus
$0.623$)---and no other term crosses the threshold in either direction. For
forecast dispersion, no term of any kind exceeds $P=0.08$ in either
specification. The tail episodes trimmed by winsorization thus do not drive
the selection.

Finally, both outcomes are persistent time series, so part of the
covariation picked up by the text terms could in principle be absorbed by
the outcome's own history. We therefore re-estimate the baseline
specification with a lag of the dependent variable added to the candidate
set. Because the survey collected in the month of the press conference may
itself postdate the conference (Section~\ref{sec:application}), a one-month
lag of the outcome would not be strictly predetermined; we therefore use the
survey collected in the month \emph{before} the press conference, that is, a
two-month lag of the outcome variable. The resulting inclusion probabilities
are reported in Table~\ref{tab:pip_ylag} in the appendix. The exercise sets
a deliberately demanding benchmark: because the outcome is a monthly series
of \emph{overlapping} fixed-horizon forecasts, consecutive surveys are
forecasts of nearly the same object, and the lag is close to a sufficient
statistic for the level. The posterior treats it accordingly: the linear lag
term is selected with near certainty ($P(\gamma=1)=1.00$) and absorbs
$\hat\pi=0.95$ of the explained variation, while no other term
survives---the monetary-policy score falls to $P=0.015$, Syuzhet to
$0.007$, and the Lagarde and COVID-19 indicators to $0.095$ and $0.081$;
for dispersion, even the own lag is not selected ($P=0.119$). The text
terms of the baseline specification thus track the persistent level of
professional expectations rather than month-to-month revisions---making
concrete the descriptive reading of the inclusion probabilities emphasized
in Section~\ref{sec:application}. Whether central bank language predicts
expectation \emph{revisions} at monthly frequency is a distinct, and
substantially harder, question that our design does not address.

\section{Implications for text-as-data practice}
\label{sec:practice}

Our application suggests four practical lessons for researchers who use
text-based covariates.

\emph{First, enter the menu, not the choice.} When several operationalizations
of a construct exist---dictionaries, embeddings, indices---the defensible
default is to let a selection-consistent procedure see all of them. In our
application, conclusions about whether ``tone matters'' would have been
reversed by a different, equally conventional, dictionary choice.

\emph{Second, treat functional form as part of the model space, even when
the answer is negative.} Every complexity and readability measure in our
application was tested for a nonlinear effect alongside a linear one, and
none showed evidence of either. A framework that only ever fits linear terms
cannot distinguish this calibrated null from the alternative that
nonlinearity was never on the table to begin with; multiverse-style
enumeration would settle the question only if a nonlinear specification had
been included in the universe by hand, and would still lack the probabilistic
calibration to say how strongly the data favor exclusion. Reporting the
linear/nonlinear decomposition for excluded terms, not only for selected
ones, is what makes a functional-form null credible rather than merely
assumed.

\emph{Third, report what was not selected---and calibrate against the prior.}
Inclusion probabilities near zero for nearly all Q\&A terms, and a dispersion
model in which no text measure exceeds an inclusion probability of
0.02---far below the prior value of one half---are substantive findings with
a precise probabilistic interpretation. The same calibration cuts the other
way: probabilities near 0.5 (such as the president's-answers term above)
represent no updating, not moderate evidence, and should be reported as
such. Reporting them guards against the file-drawer asymmetry in
which only the selected indicator ever reaches the reader.

\emph{Fourth, respect the limits.} The framework quantifies association under
model uncertainty; it does not by itself deliver causal identification, which
for text-based treatments raises distinct problems
\citep{egami2022}. Nor does it absolve the researcher of measurement care:
selection operates on the covariates as constructed, so upstream choices
\citep{denny2018} still deserve scrutiny---ideally, by including competing
constructions in the candidate set, as we do here. Finally, our
observations form a time series of press-conference events, and the additive
model treats
disturbances as independent; extending the framework with serially correlated
errors is a natural next step, and the persistence of expectations data means
inclusion probabilities should be read as measures of systematic association
rather than as tests with exact frequentist error rates. One further
robustness exercise belongs in any application of this kind: sensitivity of
the selection to the prior inclusion weight (our $\mathrm{Beta}(1,1)$ prior
implies a prior inclusion probability of one half; sparser priors would
stress-test the borderline terms). Sensitivity to the timing convention and
to winsorization we examine directly in Section~\ref{sec:robustness}.

\section{Conclusion}
\label{sec:conclusion}

Text-as-data methods have given political science an abundance of measures;
they have not, by themselves, told us which measures belong in our models.
This paper has argued that the resulting multiplicity is a model-uncertainty
problem and that it can be handled with the same Bayesian discipline the
field already applies to covariate selection---extended, crucially, to
functional form. Spike-and-slab generalized additive models deliver both
layers in one posterior: inclusion probabilities for every candidate text
measure, and a data-driven verdict on whether each selected measure acts
linearly or nonlinearly.

The ECB application shows the machinery at work on a politically consequential
corpus: the public justifications of a powerful unelected institution.
Professional forecasters, the audience with the strongest incentives to parse
those justifications, respond to a sparse and specific subset of the ECB's
language---the policy-stance tone and general-purpose sentiment of the
scripted statement---and to almost nothing else: not to its readability or
syntactic complexity, which the joint posterior tests for and does not find,
not to the unscripted exchange, and not in their disagreement, which tracks
no text measure at all. Two of the sentiment measures most widely
used in the literature carry no incremental information. None of
these conclusions---the two robust associations, the functional-form null,
and the Q\&A and dispersion nulls---could have been established, with
quantified uncertainty, by a design that pre-selects a single indicator and a
linear form.

Extensions suggest themselves in both directions. Methodologically, the
candidate set can be broadened to topic shares, embedding-based scores, and
ideal points; interactions between text measures and institutional context
are selectable terms like any other; and serially dependent error structures
would tailor the framework further to time-series corpora. Substantively,
the same design applies wherever institutions govern through justification
---courts, regulatory agencies, international organisations---and wherever
researchers must decide, without decisive theory, which features of language
to put into a model.

\paragraph*{Acknowledgments.}
The views expressed in this paper do not necessarily reflect those of the
Oesterreichische Nationalbank or the Eurosystem.
The authors used Anthropic's Claude, via the Claude Code command-line
interface (Claude Sonnet~4.5 and Claude Sonnet~5;
\url{https://claude.com/claude-code}), between January 2026 and August 2026 to
assist with writing and debugging \textsf{R} code for data preparation and
estimation, with formatting \LaTeX{} tables and figures, and with language
editing and the drafting and revision of prose. The tool was not used to
generate research findings or to analyse data: all estimation was carried out
with named \textsf{R} packages, and every reported quantity was verified by
the authors against the estimation output. The authors are entirely
responsible for the scientific content of the paper, which adheres to the
journal's authorship policy.

\paragraph*{Data availability statement.}
Replication materials---data, code, and a run log, including the MCMC
seed---will be deposited in a public data repository (the Harvard Dataverse)
(\url{https://dataverse.harvard.edu/}) upon acceptance.

\paragraph*{Financial support.}
This research received no specific grant from any funding agency, commercial
or not-for-profit sectors.

\paragraph*{Competing interests.}
The authors declare none.

\appendix
\section{Textual measures}
\label{app:measures}

\begin{table}[H]
\centering
\caption{Textual measures and variable prefixes}
\label{tab:measures}
\small
\begin{tabular}{llp{5.8cm}l}
\toprule
Dimension & Measure & Description & Prefix\\
\midrule
\multirow{4}{*}{Sentiment}
  & Syuzhet & General-purpose sentiment polarity lexicon
    \citep{Syuzhet,kim2022syuzhet} & \texttt{syuz}\\[3pt]
  & LM & Financial sentiment dictionary \citep{loughran2011};
    net polarity $(N_+-N_-)/(N_++N_-)$ & \texttt{lm}\\[3pt]
  & FinBERT & Net sentiment (positive minus negative probability) from a
    financial-domain BERT model & \texttt{finbert\_net\_sent}\\[3pt]
  & PR tone & Monetary-policy and economic-environment tone scores from the
    ECB-specific Picault--Renault lexicon \citep{picault2017} &
    \texttt{mp\_pr}, \texttt{ec\_pr}\\
\midrule
\multirow{3}{*}{Readability}
  & Flesch Ease & Flesch Reading Ease score (higher = easier); in logs & \texttt{flesch\_ease}\\[3pt]
  & FK Grade & Flesch--Kincaid grade level & \texttt{fk\_grade}\\[3pt]
  & ARI & Automated Readability Index & \texttt{ari}\\
\midrule
\multirow{3}{*}{\shortstack[l]{Complexity\\(syntactic)}}
  & Avg.\ wps & Average words per sentence; in logs & \texttt{avg\_wps}\\[3pt]
  & Complex ratio & Share of multi-clause sentences & \texttt{complex\_ratio}\\[3pt]
  & CCI & Conceptual complexity index following \citet{mcmahon2026} & \texttt{cci}\\
\midrule
\multirow{2}{*}{\shortstack[l]{Complexity\\(lexical)}}
  & Avg.\ chars & Average characters per word & \texttt{avg\_chars\_per\_word}\\[3pt]
  & TTR & Type--token ratio (distinct word types / total word tokens); mechanically
    sensitive to text length & \texttt{type\_token\_ratio}\\
\midrule
\multirow{3}{*}{Length}
  & Char count & Total characters; in logs & \texttt{char\_count}\\[3pt]
  & Sentence count & Total sentences; in logs & \texttt{sentence\_count}\\[3pt]
  & Paragraph count & Total paragraphs & \texttt{paragraph\_count}\\
\bottomrule
\multicolumn{4}{p{0.92\linewidth}}{\footnotesize\emph{Notes:} Measures are
  computed separately for the introductory statement (\texttt{intro\_}), the
  full Q\&A session (\texttt{qa\_}), and the president's answers within the
  Q\&A (\texttt{qa\_ans\_}); the Picault--Renault tone scores and the FinBERT
  score are computed for the introductory statement and the full Q\&A only.
  Sentiment measures (Syuzhet, LM, FinBERT, PR tone) are each computed once
  on the entire text of a source, not averaged over its constituent
  sentences. Several indicators enter the
  model after transformation (e.g., logarithms of length-type measures); all
  continuous indicators are winsorized at the 1st/99th percentiles. Exact
  definitions, variable names, and transformations are documented in the
  replication materials.}
\end{tabular}
\end{table}

\section{MCMC diagnostics}
\label{app:mcmc}

\begin{table}[H]
\centering
\caption{MCMC convergence and model-fit diagnostics --- one-month lead ($t+1$) specification}
\label{tab:mcmc_diag_t1}
\smallskip\par
\input{Figures/mcmc_diag_t1_19.tex}
\end{table}

\begin{table}[H]
\centering
\caption{MCMC convergence and model-fit diagnostics --- contemporaneous ($t$) specification}
\label{tab:mcmc_diag_t0}
\smallskip\par
\small
\textit{Panel A: Model fit}
\smallskip\par
\begin{tabular}{lccc}
\toprule
Model & Null deviance & Mean posterior deviance & Mass on null model \\
\midrule
Forecast level (CM\_w)       & --- & --- & $<0.001$ \\
Forecast dispersion (CS\_w)  & --- & --- & $<0.001$ \\
\bottomrule
\end{tabular}

\bigskip
\textit{Panel B: Gelman--Rubin $\hat{R}$ and effective sample size}
\smallskip\par
\begin{tabular}{llcc}
\toprule
Model & Parameter & $\hat{R}$ & ESS \\
\midrule
\multirow{5}{*}{Forecast level (CM\_w)}
  & $\sigma^2$                          & 1.009 & 34{,}897 \\
  & Log-likelihood                      & 1.000 & 23{,}229 \\
  & lin(\texttt{intro\_syuz})           & 1.000 & $\phantom{0}$8{,}726 \\
  & lin(\texttt{intro\_mp\_pr})         & 1.000 & 13{,}089 \\
  & lin(\texttt{intro\_fk\_grade})      & 1.008 & 41{,}202 \\
\midrule
\multirow{2}{*}{Forecast dispersion (CS\_w)}
  & $\sigma^2$                          & 1.000 & 48{,}869 \\
  & Log-likelihood                      & 1.000 & 48{,}566 \\
\bottomrule
\multicolumn{4}{p{0.85\linewidth}}{\footnotesize\emph{Notes:}
Panel A: null deviance $= -2\log L$ of the intercept-only model; mean
posterior deviance $= E[-2\log L \mid \mathbf{y}]$; both unavailable from the
fitted object for this specification (\texttt{---}); mass on null model
$=$ posterior probability that all inclusion indicators $\gamma_k = 0$,
estimated from MCMC draws. Panel B: $\hat{R}$ is the Gelman--Rubin
potential scale reduction factor computed across 5 parallel chains; values
below 1.1 indicate convergence. ESS is the effective sample size from the
combined chain draws (5 chains $\times$ 1{,}000 retained draws each after
burn-in and thinning). Reported parameters refer to the linear component
of the two robustly selected continuous terms in the forecast-level model
and, for comparison, the Flesch--Kincaid grade level.}
\end{tabular}
\end{table}

\section{Robustness: contemporaneous specification}
\label{app:t0}

\begin{figure}[H]
  \centering
  \begin{subfigure}[b]{0.45\textwidth}
    \centering
    \includegraphics[width=\linewidth]{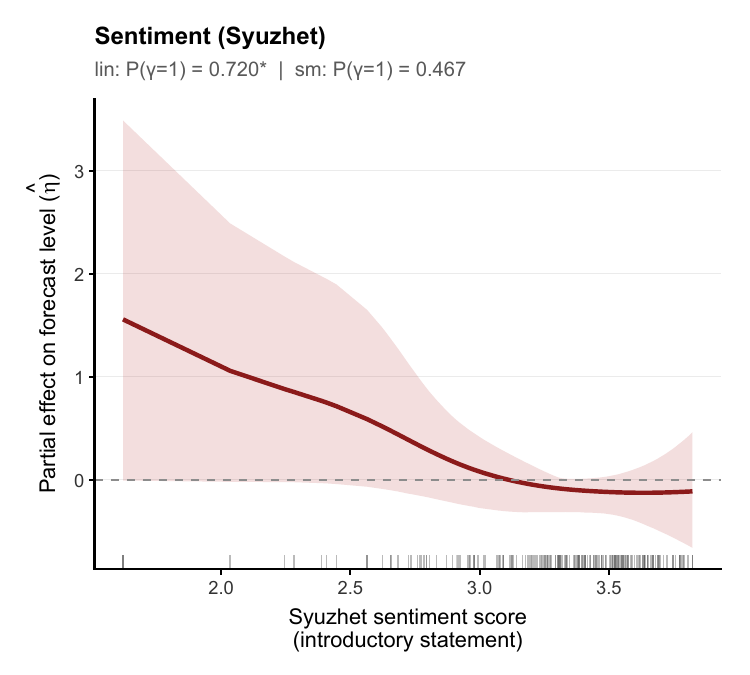}
    \subcaption{Syuzhet sentiment (\texttt{intro\_syuz})}
    \label{fig:t0_syuz}
  \end{subfigure}

  \medskip
  \begin{subfigure}[b]{0.45\textwidth}
    \centering
    \includegraphics[width=\linewidth]{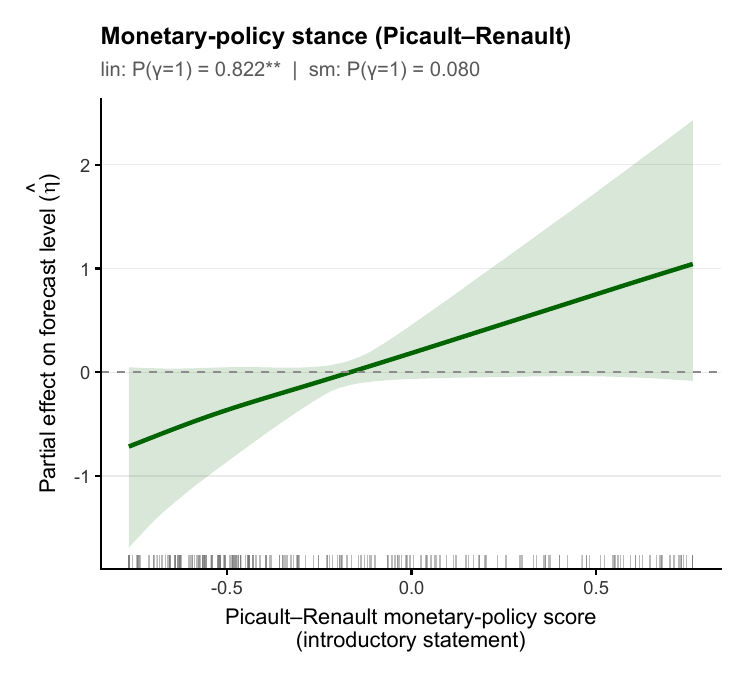}
    \subcaption{Monetary-policy stance (\texttt{intro\_mp\_pr})}
    \label{fig:t0_mp}
  \end{subfigure}

  \medskip
  \begin{subfigure}[b]{0.45\textwidth}
    \centering
    \includegraphics[width=\linewidth]{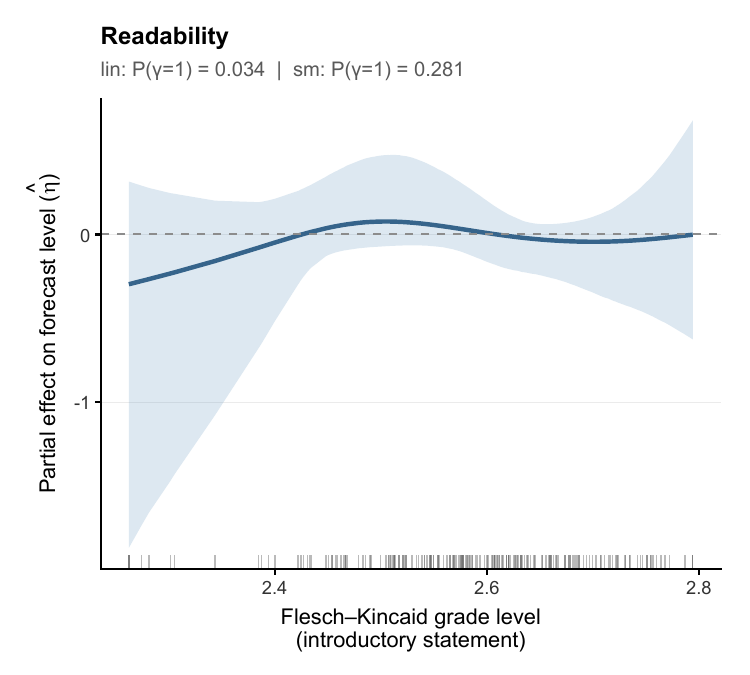}
    \subcaption{Flesch--Kincaid grade (\texttt{intro\_fk\_grade}); not selected}
    \label{fig:t0_fkg}
  \end{subfigure}

  \caption{Posterior partial effects, contemporaneous ($t$) specification
           (winsorized): panels (a)--(b) show the two selected
           introductory-statement tone measures; panel (c) shows the
           Flesch--Kincaid grade level as an illustrative example of a
           nonlinear term, although it is not selected}
  \label{fig:t0_partial}
  \smallskip\par
  \begin{minipage}{0.95\textwidth}
    \footnotesize\emph{Notes:} Posterior mean partial effect (solid line) and
    95\% credible band (5,000 MCMC draws, 5 chains). Rug marks show the
    predictor's empirical distribution. Contemporaneous ($t$) specification;
    winsorized; no macro controls.
  \end{minipage}
\end{figure}

Table~\ref{tab:pip_t0} and Figure~\ref{fig:t0_partial} report results under
the contemporaneous timing convention, in which the survey collected in the
same calendar month as the press conference serves as the outcome, rather
than the one-month lead convention used as the baseline in the main text
(Section~\ref{sec:results}). The conclusions are unchanged, and if anything
somewhat stronger for the two selected tone measures. The Syuzhet lexicon is
robustly selected in its linear component ($P(\gamma=1)=0.720$; baseline
$0.660$), entering with a negative coefficient---more optimistic
introductory-statement language is associated with lower professional CPI
inflation expectations---while its nonlinear component remains sub-threshold
($P=0.467$; baseline $0.442$). The Picault--Renault monetary-policy score is
again robustly and positively selected ($P=0.822$; baseline $0.841$),
confirming that a more hawkish introductory statement is associated with
higher inflation expectations irrespective of the timing convention, and
remains the single most robustly selected text term under both
specifications. The functional-form null carries over without exception:
the Flesch--Kincaid nonlinear component remains sub-threshold
($P=0.281$; baseline $0.315$), as does every other complexity and
readability measure. The type--token ratio of the president's Q\&A answers,
the closest Q\&A contender, is sub-threshold under both timing conventions
($P=0.460$; baseline $0.429$)---the two specifications agree on this term as
on every other. The Lagarde presidency term is again the most strongly
selected categorical control ($P=0.930$; baseline $0.946$), and the
COVID-19 crisis indicator is also selected in this specification
($P=0.602$; baseline $0.623$). For forecast disagreement the null
result carries over without exception: no text-based term exceeds
$P(\gamma=1)=0.01$ in the contemporaneous specification.

\begin{table}[H]
\centering
\caption{Posterior inclusion probabilities, posterior mean coefficients, and term importance --- contemporaneous ($t$) specification}
\label{tab:pip_t0}
\smallskip\par
\input{Figures/pip_table_main_18.tex}
\end{table}

\section{Robustness: non-winsorized specification}
\label{app:nonwinsor}

Table~\ref{tab:pip_nonwinsor} reports posterior inclusion probabilities and
term importances for the baseline one-month lead ($t+1$) specification
re-estimated on non-winsorized covariates, discussed in
Section~\ref{sec:robustness}.

\begin{table}[H]
\centering
\caption{Posterior inclusion probabilities and term importance --- non-winsorized one-month lead ($t+1$) specification}
\label{tab:pip_nonwinsor}
\smallskip\par
\small
\input{Figures/pip_table_t1_nonwinsor_20.tex}
\end{table}

\section{Robustness: lagged dependent variable}
\label{app:ylag}

Table~\ref{tab:pip_ylag} reports posterior inclusion probabilities and term
importances for the baseline one-month lead ($t+1$, winsorized)
specification with a two-month lag of the dependent variable---the survey
collected in the month before the press conference---added to the candidate
set, discussed in Section~\ref{sec:robustness}.

\begin{table}[H]
\centering
\caption{Posterior inclusion probabilities and term importance --- one-month lead ($t+1$) specification with lagged dependent variable}
\label{tab:pip_ylag}
\smallskip\par
\small
\input{Figures/pip_table_t1_ylag_23.tex}
\end{table}

\printbibliography

\end{document}

%% file: Figures/pip_table_t1_18.tex
\begin{tabular}{lcccccc}
\toprule
& \multicolumn{3}{c}{Forecast Level ($\pi^e_{t+1}$)} & \multicolumn{3}{c}{Forecast Dispersion ($\sigma(\pi^e_{t+1})$)}\\
\cmidrule(lr){2-4}\cmidrule(lr){5-7}
Variable & $P(\gamma=1)$ & $\hat\beta$ & $\hat\pi$ & $P(\gamma=1)$ & $\hat\beta$ & $\hat\pi$\\
\midrule
\multicolumn{7}{l}{\textit{Introductory statement: sentiment}}\\
\quad lin(\texttt{intro\_syuz}) & 0.660$^{*}$ & $-0.443$ & $\phantom{-}0.15$ & 0.000 & $\phantom{-}0.000$ & $-0.00$\\
\quad sm(\texttt{intro\_syuz}) & 0.442 & --- & $\phantom{-}0.03$ & 0.001 & --- & $\phantom{-}0.02$\\
\quad lin(\texttt{intro\_lm}) & 0.029 & $-0.015$ & $\phantom{-}0.00$ & 0.003 & $-0.059$ & $\phantom{-}0.04$\\
\quad lin(\texttt{finbert\_net\_sentiment\_intro}) & 0.020 & $\phantom{-}0.004$ & $-0.00$ & 0.000 & $-0.009$ & $\phantom{-}0.00$\\
\midrule
\multicolumn{7}{l}{\textit{Introductory statement: Picault--Renault tone}}\\
\quad lin(\texttt{intro\_mp\_pr}) & 0.841$^{**}$ & $\phantom{-}1.107$ & $\phantom{-}0.18$ & 0.003 & $\phantom{-}0.025$ & $\phantom{-}0.04$\\
\quad lin(\texttt{intro\_ec\_pr}) & 0.034 & $\phantom{-}0.018$ & $\phantom{-}0.00$ & 0.001 & $\phantom{-}0.011$ & $\phantom{-}0.01$\\
\midrule
\multicolumn{7}{l}{\textit{Introductory statement: complexity}}\\
\quad sm(\texttt{intro\_fk\_grade}) & 0.315 & --- & $-0.01$ & 0.001 & --- & $-0.00$\\
\quad lin(\texttt{intro\_complex\_ratio}) & 0.236 & $-0.162$ & $\phantom{-}0.05$ & 0.003 & $-0.042$ & $\phantom{-}0.02$\\
\quad sm(\texttt{intro\_avg\_wps}) & 0.257 & --- & $-0.00$ & 0.000 & --- & $-0.00$\\
\quad lin(\texttt{intro\_avg\_chars\_per\_word}) & 0.094 & $\phantom{-}0.049$ & $-0.01$ & 0.000 & $\phantom{-}0.005$ & $-0.00$\\
\midrule
\multicolumn{7}{l}{\textit{Q\&A session}}\\
\quad lin(\texttt{qa\_ans\_type\_token\_ratio}) & 0.429 & $\phantom{-}0.275$ & $\phantom{-}0.03$ & 0.001 & $\phantom{-}0.006$ & $\phantom{-}0.00$\\
\quad lin(\texttt{qa\_mp\_pr}) & 0.171 & $\phantom{-}0.149$ & $\phantom{-}0.03$ & 0.002 & $\phantom{-}0.023$ & $\phantom{-}0.04$\\
\quad All remaining Q\&A terms & $<$0.30 & --- & $\approx$0 & $<$0.02 & --- & $\approx$0\\
\midrule
\multicolumn{7}{l}{\textit{Structural indicators}}\\
\quad fct(\texttt{lagarde}) & 0.946$^{***}$ & --- & $\phantom{-}0.50$ & 0.009 & --- & $-0.01$\\
\quad fct(\texttt{krisecorona}) & 0.623$^{*}$ & --- & $\phantom{-}0.03$ & 0.044 & --- & $-0.02$\\
\quad fct(\texttt{draghi}) & 0.081 & --- & $-0.01$ & 0.028 & --- & $\phantom{-}0.20$\\
\quad fct(\texttt{trichet}) & 0.039 & --- & $-0.00$ & 0.003 & --- & $\phantom{-}0.02$\\
\bottomrule
\multicolumn{7}{p{0.95\linewidth}}{\footnotesize\emph{Notes:}
$P(\gamma=1)$: marginal posterior inclusion probability under the peNMIG prior.
$\hat\beta$: posterior mean coefficient for \texttt{lin} and \texttt{fct} terms
(Frobenius-normalised scale, comparable across terms); \texttt{---} for
nonlinear smooth terms --- see partial-effects figures for direction.
$\hat\pi_k$: term importance (equation~\eqref{eq:pi}); magnitudes indicate
relative importance; can be negative for correlated terms.
$^{*}P(\gamma=1)>0.5$; $^{**}P(\gamma=1)>0.75$; $^{***}P(\gamma=1)>0.9$.
Winsorized specification; no macro controls. CE forecasts matched to press-conference month $t+1$.
}
\end{tabular}

%% file: Figures/single_indicator_ols_22.tex
\begin{tabular}{lcc}
\toprule
Indicator ($y=\pi^e_{t+1}$) & Bivariate & With controls\\
\midrule
Syuzhet & -1.217$^{***}$ (0.164) & -0.419$^{**}$ (0.195)\\
Loughran--McDonald & -0.799$^{***}$ (0.167) & -0.134 (0.182)\\
FinBERT & -0.967$^{***}$ (0.288) & 0.268 (0.269)\\
PR monetary-policy & 0.501$^{***}$ (0.154) & 1.570$^{***}$ (0.192)\\
PR economic-environment & 0.468 (0.298) & 0.647$^{**}$ (0.270)\\
FK grade (linear) & -2.484$^{***}$ (0.641) & 0.920 (0.737)\\
\bottomrule
\multicolumn{3}{p{0.85\linewidth}}{\footnotesize\emph{Notes:}
Each cell reports the coefficient on the indicator from a separate OLS
regression of the one-month-lead ($t+1$) forecast level on that indicator
alone (bivariate) or with president and crisis-period dummies (controls).
OLS standard errors in parentheses; disturbances are treated as
independent, as in the SSGAM specification.
$^{*}$/$^{**}$/$^{***}$: $p<0.10$/$0.05$/$0.01$. Covariates winsorized at
the 1st/99th percentiles and sample restricted by listwise deletion across
the full covariate menu, matching the SSGAM baseline
($n=224$).}
\end{tabular}

%% file: Figures/mcmc_diag_t1_19.tex
\small
\textit{Panel A: Model fit}
\smallskip\par
\begin{tabular}{lccc}
\toprule
Model & Null deviance & Mean posterior deviance & Mass on null model \\
\midrule
Forecast level (CM\_w)       & --- & --- & $<0.001$ \\
Forecast dispersion (CS\_w)       & --- & --- & $<0.001$ \\
\bottomrule
\end{tabular}

\bigskip
\textit{Panel B: Gelman--Rubin $\hat{R}$ and effective sample size}
\smallskip\par
\begin{tabular}{llcc}
\toprule
Model & Parameter & $\hat{R}$ & ESS \\
\midrule
\multirow{5}{*}{Forecast level (CM\_w)}
  & $\sigma^2$                          & 1.007 & 17{,}521 \\
  & Log-likelihood                      & 1.000 & 24{,}521 \\
  & lin(\texttt{intro\_syuz})      & 1.001 & 8{,}773 \\
  & lin(\texttt{intro\_mp\_pr})      & 1.000 & 14{,}706 \\
  & lin(\texttt{intro\_fk\_grade})      & 1.002 & 34{,}896 \\
\midrule
\multirow{2}{*}{Forecast dispersion (CS\_w)}
  & $\sigma^2$                          & 1.068 & 18{,}957 \\
  & Log-likelihood                      & 1.000 & 33{,}863 \\
\bottomrule
\multicolumn{4}{p{0.85\linewidth}}{\footnotesize\emph{Notes:}
Panel A: null deviance $= -2\log L$ of the intercept-only model; mean
posterior deviance $= E[-2\log L \mid \mathbf{y}]$; mass on null model
$=$ posterior probability that all inclusion indicators $\gamma_k = 0$,
estimated from MCMC draws. Panel B: $\hat{R}$ is the Gelman--Rubin
potential scale reduction factor computed across 5 parallel chains; values
below 1.1 indicate convergence. ESS is the effective sample size from the
combined chain draws. Selected beta parameters refer to the linear component
of the three robustly selected continuous terms in the forecast-level model.
One-month lead ($t+1$) specification.}
\end{tabular}

%% file: Figures/pip_table_main_18.tex
\begin{tabular}{lcccccc}
\toprule
& \multicolumn{3}{c}{Forecast Level ($\pi^e_t$)} & \multicolumn{3}{c}{Forecast Dispersion ($\sigma(\pi^e_t)$)}\\
\cmidrule(lr){2-4}\cmidrule(lr){5-7}
Variable & $P(\gamma=1)$ & $\hat\beta$ & $\hat\pi$ & $P(\gamma=1)$ & $\hat\beta$ & $\hat\pi$\\
\midrule
\multicolumn{7}{l}{\textit{Introductory statement: sentiment}}\\
\quad lin(\texttt{intro\_syuz}) & 0.720$^{*}$ & $-0.494$ & $\phantom{-}0.18$ & 0.000 & $-0.010$ & $\phantom{-}0.01$\\
\quad sm(\texttt{intro\_syuz}) & 0.467 & --- & $\phantom{-}0.04$ & 0.001 & --- & $\phantom{-}0.01$\\
\quad lin(\texttt{intro\_lm}) & 0.023 & $-0.010$ & $\phantom{-}0.00$ & 0.001 & $-0.037$ & $\phantom{-}0.03$\\
\quad lin(\texttt{finbert\_net\_sentiment\_intro}) & 0.025 & $-0.010$ & $\phantom{-}0.00$ & 0.001 & $-0.012$ & $\phantom{-}0.00$\\
\midrule
\multicolumn{7}{l}{\textit{Introductory statement: Picault--Renault tone}}\\
\quad lin(\texttt{intro\_mp\_pr}) & 0.822$^{**}$ & $\phantom{-}1.019$ & $\phantom{-}0.16$ & 0.007 & $\phantom{-}0.039$ & $\phantom{-}0.06$\\
\quad lin(\texttt{intro\_ec\_pr}) & 0.027 & $\phantom{-}0.010$ & $\phantom{-}0.00$ & 0.002 & $\phantom{-}0.025$ & $\phantom{-}0.03$\\
\midrule
\multicolumn{7}{l}{\textit{Introductory statement: complexity}}\\
\quad sm(\texttt{intro\_fk\_grade}) & 0.281 & --- & $-0.01$ & 0.001 & --- & $-0.00$\\
\quad lin(\texttt{intro\_complex\_ratio}) & 0.393 & $-0.345$ & $\phantom{-}0.12$ & 0.004 & $-0.050$ & $\phantom{-}0.04$\\
\quad sm(\texttt{intro\_avg\_wps}) & 0.194 & --- & $-0.01$ & 0.000 & --- & $-0.00$\\
\quad lin(\texttt{intro\_avg\_chars\_per\_word}) & 0.298 & $\phantom{-}0.202$ & $-0.03$ & 0.000 & $\phantom{-}0.002$ & $-0.00$\\
\midrule
\multicolumn{7}{l}{\textit{Q\&A session}}\\
\quad lin(\texttt{qa\_ans\_type\_token\_ratio}) & 0.460 & $\phantom{-}0.300$ & $\phantom{-}0.04$ & 0.001 & $\phantom{-}0.018$ & $\phantom{-}0.01$\\
\quad lin(\texttt{qa\_mp\_pr}) & 0.159 & $\phantom{-}0.123$ & $\phantom{-}0.03$ & 0.003 & $\phantom{-}0.028$ & $\phantom{-}0.05$\\
\quad All remaining Q\&A terms & $<$0.23 & --- & $\approx$0 & $<$0.01 & --- & $\approx$0\\
\midrule
\multicolumn{7}{l}{\textit{Structural indicators}}\\
\quad fct(\texttt{lagarde}) & 0.930$^{***}$ & --- & $\phantom{-}0.44$ & 0.010 & --- & $-0.02$\\
\quad fct(\texttt{krisecorona}) & 0.602$^{*}$ & --- & $\phantom{-}0.02$ & 0.003 & --- & $-0.01$\\
\quad fct(\texttt{draghi}) & 0.080 & --- & $-0.01$ & 0.017 & --- & $\phantom{-}0.16$\\
\quad fct(\texttt{trichet}) & 0.038 & --- & $-0.00$ & 0.006 & --- & $\phantom{-}0.04$\\
\bottomrule
\multicolumn{7}{p{0.95\linewidth}}{\footnotesize\emph{Notes:}
$P(\gamma=1)$: marginal posterior inclusion probability under the peNMIG prior.
$\hat\beta$: posterior mean coefficient for \texttt{lin} and \texttt{fct} terms
(Frobenius-normalised scale, comparable across terms); \texttt{---} for
nonlinear smooth terms --- see partial-effects figures for direction.
$\hat\pi_k$: term importance (equation~\eqref{eq:pi}); magnitudes indicate
relative importance; can be negative for correlated terms.
$^{*}P(\gamma=1)>0.5$; $^{**}P(\gamma=1)>0.75$; $^{***}P(\gamma=1)>0.9$.
Winsorized specification; no macro controls.
}
\end{tabular}

%% file: Figures/pip_table_t1_nonwinsor_20.tex
\begin{tabular}{lcccc}
\toprule
& \multicolumn{2}{c}{Forecast Level ($\pi^e_{t+1}$)} & \multicolumn{2}{c}{Forecast Dispersion ($\sigma(\pi^e_{t+1})$)}\\
\cmidrule(lr){2-3}\cmidrule(lr){4-5}
Variable & $P(\gamma=1)$ & $\hat\pi$ & $P(\gamma=1)$ & $\hat\pi$\\
\midrule
\multicolumn{5}{l}{\textit{Introductory statement: sentiment}}\\
\quad lin(\texttt{intro\_syuz}) & 0.739$^{*}$ & $\phantom{-}$0.17 & 0.000 & $\phantom{-}$0.00\\
\quad sm(\texttt{intro\_syuz}) & 0.309 & $\phantom{-}$0.02 & 0.002 & $\phantom{-}$0.02\\
\quad lin(\texttt{intro\_lm}) & 0.020 & $\phantom{-}$0.00 & 0.003 & $\phantom{-}$0.04\\
\quad lin(\texttt{finbert\_net\_sentiment\_intro}) & 0.021 & $-$0.00 & 0.001 & $\phantom{-}$0.00\\
\midrule
\multicolumn{5}{l}{\textit{Introductory statement: Picault--Renault tone}}\\
\quad lin(\texttt{intro\_mp\_pr}) & 0.846$^{**}$ & $\phantom{-}$0.18 & 0.003 & $\phantom{-}$0.04\\
\quad lin(\texttt{intro\_ec\_pr}) & 0.033 & $\phantom{-}$0.00 & 0.001 & $\phantom{-}$0.02\\
\midrule
\multicolumn{5}{l}{\textit{Introductory statement: complexity}}\\
\quad sm(\texttt{intro\_fk\_grade}) & 0.310 & $-$0.01 & 0.001 & $-$0.00\\
\quad lin(\texttt{intro\_complex\_ratio}) & 0.215 & $\phantom{-}$0.04 & 0.004 & $\phantom{-}$0.03\\
\quad sm(\texttt{intro\_avg\_wps}) & 0.217 & $-$0.00 & 0.000 & $-$0.00\\
\quad lin(\texttt{intro\_avg\_chars\_per\_word}) & 0.051 & $-$0.00 & 0.001 & $-$0.00\\
\midrule
\multicolumn{5}{l}{\textit{Q\&A session}}\\
\quad lin(\texttt{qa\_ans\_type\_token\_ratio}) & 0.446 & $\phantom{-}$0.04 & 0.001 & $\phantom{-}$0.01\\
\quad lin(\texttt{qa\_mp\_pr}) & 0.175 & $\phantom{-}$0.03 & 0.002 & $\phantom{-}$0.05\\
\quad All remaining Q\&A terms & $<$0.30 & $\approx$0 & $<$0.03 & $\approx$0\\
\midrule
\multicolumn{5}{l}{\textit{Structural indicators}}\\
\quad fct(\texttt{lagarde}) & 0.953$^{***}$ & $\phantom{-}$0.49 & 0.005 & $-$0.00\\
\quad fct(\texttt{krisecorona}) & 0.703$^{*}$ & $\phantom{-}$0.04 & 0.007 & $-$0.01\\
\quad fct(\texttt{draghi}) & 0.075 & $-$0.01 & 0.033 & $\phantom{-}$0.21\\
\quad fct(\texttt{trichet}) & 0.036 & $-$0.00 & 0.002 & $\phantom{-}$0.02\\
\bottomrule
\multicolumn{5}{p{0.95\linewidth}}{\footnotesize\emph{Notes:}
$P(\gamma=1)$: marginal posterior inclusion probability.
$\hat\pi_k$: term importance (equation~\eqref{eq:pi}); magnitudes indicate
relative importance. $^{*}$: $P(\gamma=1)>0.5$; $^{**}$: $P(\gamma=1)>0.75$;
$^{***}$: $P(\gamma=1)>0.9$. Non-winsorized specification, no macro controls.
CE forecasts matched to press-conference month $t+1$ (one month lead).}
\end{tabular}

%% file: Figures/pip_table_t1_ylag_23.tex
\begin{tabular}{lcccc}
\toprule
& \multicolumn{2}{c}{Forecast Level ($\pi^e_{t+1}$)} & \multicolumn{2}{c}{Forecast Dispersion ($\sigma(\pi^e_{t+1})$)}\\
\cmidrule(lr){2-3}\cmidrule(lr){4-5}
Variable & $P(\gamma=1)$ & $\hat\pi$ & $P(\gamma=1)$ & $\hat\pi$\\
\midrule
\multicolumn{5}{l}{\textit{Lagged dependent variable}}\\
\quad lin(\texttt{y\_lag2}) & 1.000$^{***}$ & $\phantom{-}$0.95 & 0.119 & $\phantom{-}$0.54\\
\quad sm(\texttt{y\_lag2}) & 0.006 & $\phantom{-}$0.00 & 0.001 & $\phantom{-}$0.01\\
\midrule
\multicolumn{5}{l}{\textit{Introductory statement: sentiment}}\\
\quad lin(\texttt{intro\_syuz}) & 0.007 & $\phantom{-}$0.01 & 0.000 & $-$0.01\\
\quad sm(\texttt{intro\_syuz}) & 0.005 & $\phantom{-}$0.00 & 0.001 & $\phantom{-}$0.01\\
\quad lin(\texttt{intro\_lm}) & 0.006 & $\phantom{-}$0.00 & 0.001 & $\phantom{-}$0.02\\
\quad lin(\texttt{finbert\_net\_sentiment\_intro}) & 0.096 & $-$0.01 & 0.000 & $\phantom{-}$0.00\\
\midrule
\multicolumn{5}{l}{\textit{Introductory statement: Picault--Renault tone}}\\
\quad lin(\texttt{intro\_mp\_pr}) & 0.015 & $\phantom{-}$0.00 & 0.002 & $-$0.00\\
\quad lin(\texttt{intro\_ec\_pr}) & 0.012 & $\phantom{-}$0.00 & 0.001 & $\phantom{-}$0.01\\
\midrule
\multicolumn{5}{l}{\textit{Introductory statement: complexity}}\\
\quad sm(\texttt{intro\_fk\_grade}) & 0.003 & $-$0.00 & 0.001 & $-$0.00\\
\quad lin(\texttt{intro\_complex\_ratio}) & 0.003 & $\phantom{-}$0.00 & 0.001 & $\phantom{-}$0.02\\
\quad sm(\texttt{intro\_avg\_wps}) & 0.003 & $-$0.00 & 0.000 & $-$0.00\\
\quad lin(\texttt{intro\_avg\_chars\_per\_word}) & 0.002 & $\phantom{-}$0.00 & 0.000 & $-$0.00\\
\midrule
\multicolumn{5}{l}{\textit{Q\&A session}}\\
\quad lin(\texttt{qa\_ans\_type\_token\_ratio}) & 0.008 & $\phantom{-}$0.00 & 0.001 & $\phantom{-}$0.00\\
\quad lin(\texttt{qa\_mp\_pr}) & 0.005 & $\phantom{-}$0.00 & 0.001 & $\phantom{-}$0.02\\
\quad All remaining Q\&A terms & $<$0.09 & $\approx$0 & $<$0.02 & $\approx$0\\
\midrule
\multicolumn{5}{l}{\textit{Structural indicators}}\\
\quad fct(\texttt{lagarde}) & 0.095 & $\phantom{-}$0.02 & 0.005 & $-$0.00\\
\quad fct(\texttt{krisecorona}) & 0.081 & $\phantom{-}$0.00 & 0.034 & $-$0.01\\
\quad fct(\texttt{draghi}) & 0.018 & $\phantom{-}$0.01 & 0.007 & $\phantom{-}$0.08\\
\quad fct(\texttt{trichet}) & 0.008 & $-$0.00 & 0.002 & $\phantom{-}$0.02\\
\bottomrule
\multicolumn{5}{p{0.95\linewidth}}{\footnotesize\emph{Notes:}
$P(\gamma=1)$: marginal posterior inclusion probability.
$\hat\pi_k$: term importance (equation~\eqref{eq:pi}); magnitudes indicate
relative importance. $^{*}$: $P(\gamma=1)>0.5$; $^{**}$: $P(\gamma=1)>0.75$;
$^{***}$: $P(\gamma=1)>0.9$. Winsorized specification, no macro controls.
CE forecasts matched to press-conference month $t+1$ (one month lead).
\texttt{y\_lag2} is the two-month lag of the dependent variable
($\pi^e_{t-1}$), i.e.\ the survey collected in the month before the press
conference, entering as a candidate term (linear and smooth) like every
other continuous covariate.}
\end{tabular}